\documentstyle[12pt]{article}
\author{Bambi Hu$^{1,2}$ and Jian Zhou$^{1,3}$}    
\title{\large{\bf{A Frenkel-Kontorova Model with Two Spring Constants:
New Phases and Phase Transitions}}} 
    
\begin{document}    
\date{}    
\maketitle    
\begin{abstract}A discrete Frenkel-Kontorova model with two 
alternate  spring constants $k_{1}$ and $k_{2}$ is studied. It is 
found that this model has many surprising behaviours. The continuum 
version of this model is different from the sine-Gordon equation, 
the continuum version of the standard FK model.
More interestingly, it has an unpinned commensurate phase 
which is translationally invariant. A phase transition takes plae on a 
lattice point in the parameter space. We have also predicted another
interesting phase transition in which an unlocked commensurate phase breaking 
translational invariance will take place as the 
strength of the substrate potential is increased. 
\end{abstract}    

 
\vspace{1cm}
\begin{enumerate}
\item Department of Physics and Centre for Nonlinear and Complex 
Systems, Hong Kong Baptist University, Kowloon Tong, Hong Kong
\item Department of Physics, University of Houston, Houston, TX 
77204, USA
\item Institute of Applied Physics and Computational Mathematics, 
P.O.Box 8009, 100088 Beijing, P.R.China    
\end{enumerate}
\baselineskip=1.2\baselineskip     
    
\newpage    
    
\hspace{0.5cm}Many years ago Frenkel and Kontorova \cite{fk} 
proposed a discrete version of the sine-Gordon equation, which is now 
known as the Frenkel-Kontorova (FK) model. It has been widely studied 
as a model for many solid-state systems \cite{nabarro, sy, wm, bm} such 
as crystal dislocations, absorbed epitaxial monolayers and 
incommensurate structures, etc. This simple classical one-dimensional 
model consists of a group of particles connected by springs subject to an 
external sinusoidal potential. It has a rich phase diagram \cite{bm, pa}. 
Its commensurate phase is separated by infinitesimal gaps of 
incommensurate structure. In a commensurate phase, translational 
invariance is broken and no zero-frequency phonon mode exits. 
However, an incommensurate phase is translational invariant and has a 
zero-frequency phonon mode. There is a ``transition by breaking of 
analyticity" \cite{pa} for an incommensurate phase as the strength of 
substrate potential is increased. However, there is no symmetry breaking 
in this transition.    
    
\hspace{0.5cm}Many generalized FK models \cite{isihara, cf, sh, 
lh, bm1} have been studied for different purposes. In this paper, we 
propose to study the following generalised FK model,
  
\begin{equation}  
U=\frac{1} {4} \sum_nf(n,k_1,k_2)(X_{n+1}-X_n-f(n,a,b))^2 -
\frac{Vc} 
{2\pi} \sum_ncos(\frac{2\pi} {c} X_n). 
\label{energyeq}
\end{equation}  
Here $f(n,l,m)= l(1-(-1)^n)-m(1+(-1)^n)$. $X_n$ is the position of 
the $n$th particle. $a$ and $b$ are the equilibrium lengths of the two 
springs with spring constants $k_1$ and $k_2$. They satisfy the 
balance equation $k_1a=k_2b$ that is the ground state solution of
 Eq.~(\ref{energyeq}) with $ V=0$. $c$ is the period of the 
sinusoidal substrate potential. The system consists of particles 
connected alternately by two kinds of springs. Obviously, in the absence 
of a substrate potential the ground state is a dimerizated state, as 
shown in Fig. 1. In many real quasi-one-dimensional systems \cite{isihara} 
of solid state physics, such dimerization often occurs due to lattice 
distortion as in the case of Peierls instability. 

 
\hspace{0.5cm}Differentiating the potential function (1) with 
respect to $X_n$, we obtain the equations for the equilibrium 
states:

\begin{equation}
\begin{array}{l} 
k_1X_{n+1}-(k_1+k_2)X_n+k_2X_{n-1}-Vsin(\frac {2\pi} {c} 
X_n)=0\\\\
k_2X_{n+2}-(k_1+k_2)X_{n+1}+k_1X_n-Vsin(\frac {2\pi} {c} 
X_{n+1})=0  
\label{equilibriumeq1}
\end{array}
\end{equation} 
We rewrite them as 

\begin{equation} 
\begin{array}{r}
 (X_{n+1}-2 X_n+X_{n-1}) +(2g-1)  (X_{n+1}-X_n)- \\(X_n-X_{n-
1})-\frac{2V}{k_2}sin(\frac {2\pi} {c} X_n)=0\\\\
(X_{n+2}-2 X_{n+1}+X_n) +(2g^{-1}-1) (X_{n+2}-X_{n+1})-
\\(X_{n+1}-X_n)-\frac{2V}{k_1}sin(\frac {2\pi} {c} X_{n+1})=0
\label{equilibriumeq2}
\end{array}
\end{equation} 
Here $g=k_1/k_2$. We introduce fields $X$ for a particle 
with $n=even$ and $Y$ for a particle with $n=odd$. In this way, the 
continuum version of Eq.~(\ref{equilibriumeq2}) can be written as,

\begin{equation}
\begin{array}{l}
\frac{d^2X} {d^2n} +(2g-1)\frac {dX} {dn}-\frac {dY}{dn} =\frac 
{2V}{k_2}sin(\frac {2\pi} {c} X)\\\\
\frac {d^2Y} {d^2n} +(2g^{-1}-1)\frac {dY} {dn}-\frac {dX}{dn} 
=\frac {2V}{k_1} sin(\frac {2\pi} {c} Y)
\label{continueq}
\end{array}
\end{equation}
They are two coupled nonlinear equations. When $g=1$, 
Eq.~(\ref{continueq}) becomes the sine-Gordon equation \cite{bm}. Intuitively, 
we expect our model to behave like the FK model when $k_1-k_2$ is 
small. However, we find this is not always the case and we will 
discuss this point in  later sections.

From previous studies \cite{bm, pa, cf}, we know the 
locked and unlocked phases of the FK model are related to the 
distribution of the particles in the external potential wells. For a 
commensurate phase \cite{cf}, the particles are symmetrically placed 
about the bottom of a well or the top of a well, i.e., there is a 
symmetric point. It is just because of this symmetric point that a 
commensurate phase is pinned to the substrate.  To reach another 
degenerate ground state of the system, it must move its symmetric 
point over the top of a well to the other side of the well. It has to 
overcome an energy barrier; therefore, the system chooses to be 
locked to the substrate. In contrast to a commensurate phase, an 
incommensurate phase does not have such a symmetric point. 
Therefore, an incommensurate phase is unlocked and is 
translationally invariant. In this sense, we can regard an 
incommensurate phase as the phase that breaks the symmetry of a 
commensurate phase. Obviously, the nonuniformity of the spring 
constants in our model also has the same effect. It makes all the 
phases less symmetric. Intuitively an incommensurate phase of our 
model is more asymmetric than that of the FK model. So a stronger 
substrate potential is expected to lock our model. Besides, due to the 
nonuniformity, a commensurate phase possibly does not have a 
symmetric point because it only can exist in some special cases. This 
means our model could have a new commensurate phase which is 
translationally invariant and unpinned. Now we prove it in a more 
rigorous manner by a perturbative study \cite{cf}.
 
For a small amplitude $V$ of a substrate potential, we can view it as a 
perturbation to a uniform ground state of $X_n-X_{n-1}=a$ or $b$. 
$V$ is treated as the perturbation parameter. Then $X_n$ can be 
rewritten as 

\begin{equation}
X_n=E_n+\delta_n+\theta
\label{pfactoreq}
\end{equation}
Here $E_n$ is the position of the $n$th particle when $V=0$, $\theta$ 
is an overall phase shift factor and $\delta_n=0$ is equivalent to $V=0$. 
Putting Eq.~(\ref{pfactoreq}) into Eq.~(\ref{equilibriumeq1}), we 
have

\begin{equation}
\begin{array}{l}
k_2\delta_{n}-(k_1+k_2)\delta_{n-1}+k_1\delta_{n-2}-Vsin[\frac {2\pi} {c} 
(E_{n-1}+\delta_{n-1}+\theta)]=0\\\\
k_1\delta_{n+1}-(k_1+k_2)\delta_n+k_2\delta_{n-1}-Vsin[\frac 
{2\pi} {c} (E_n+\delta_n+\theta)]=0\\\\
k_2\delta_{n+2}-(k_1+k_2)\delta_{n+1}+k_1\delta_{n}-Vsin[\frac 
{2\pi} {c} (E_{n+1}+\delta_{n+1}+\theta)]=0
\label{pertubationeq1}
\end{array}
\end{equation}
Expanding $\delta_n$ in orders of $V$, we have 
$\delta_n=\delta_n^1(V)+ 
\delta_n^2(V^2)+ \delta_n^3(V^3)+....$ To first order in $V$, we get 

\begin{equation}
\begin{array}{l}
k_2\delta_{n}^1-(k_1+k_2)\delta_{n-1}^1+k_1\delta_{n-2}^1-
Vsin[\frac {2\pi} {c} (E_{n-1}+\theta)]=0\\\\
k_1\delta_{n+1}^1-(k_1+k_2)\delta_n^1+k_2\delta_{n-1}^1-
Vsin[\frac {2\pi}{c} (E_n+\theta)]=0\\\\
k_2\delta_{n+2}^1-(k_1+k_2)\delta_{n+1}^1+k_1\delta_{n}^1-
Vsin[\frac {2\pi} {c} (E_{n+1}+\theta)]=0
\label{pertubationeq2}
\end{array}
\end{equation}
Eliminating $\delta_{n+1}^1$ and $\delta_{n-1}^1$, we have
\begin{equation}
\delta_{n+2}^1-2\delta_n^1+\delta_{n-2}^1=Asin[\frac {2\pi} {c} 
(E_n+\theta)]+Bcos[\frac {2\pi} {c} (E_n+\theta)]
\label{pertubationeq3}
\end{equation}
where

$$
\begin{array}{l}
 A=\frac {V} {k_1k_2} [k_1cos(\frac {2\pi} {c}a)+k_2cos(\frac 
{2\pi} {c} b)+k_1+k_2]\\\\
B=\frac {V} {k_1k_2} [k_1sin(\frac {2\pi} {c} a)-k_2sin(\frac 
{2\pi} {c} b)]
\end{array}
$$
To obtain these results, we have made use of  $E_{n-1}=E_n-b$ and 
$E_{n+1}=E_n+a$. The solution to Eq.~(\ref{pertubationeq3}) is

\begin{equation}
\delta_n^1=\frac {Asin[\frac {2\pi} {c} (E_n+\theta)]+(-1)^n 
Bcos[\frac {2\pi} {c} (E_n+\theta)]} {2(cos[\frac {2\pi} {c} (a+b)]-
1)}.
\label{psolution1}
\end{equation}
Putting this solution into Eq.~(\ref{pertubationeq1}), we 
obtain

\begin{equation}
\delta_n^2=\frac {A_1sin[\frac {4\pi} {c} (E_n+\theta)]+(-1)^n 
B_1cos[\frac {4\pi}{c} (E_n+\theta)]} {8(cos[\frac {2\pi} {c} (a+b)]-
1)(cos[\frac {4\pi} {c}(a+b)]-1)}
\label{psolution2}
\end{equation}
with

 $$
\begin{array}{l}
 A_1=\frac {2\pi V} {k_1k_2c} [B(k_1sin(\frac {4\pi} {c} a)-
k_2sin(\frac {4\pi} {c} b))+A (k_1+k_2+k_1cos(\frac {4\pi} 
{c}a)+k_2cos(\frac {4\pi} {c} b))]\\\\
B_1=\frac {2\pi V} {k_1k_2c}[A(k_1sin(\frac {4\pi} {c} a)-
k_2sin(\frac {4\pi} {c} b))+B(k_1+k_2-k_1cos(\frac {4\pi} {c}a)-
k_2cos(\frac {4\pi} {c} b))] 
\end{array}
$$
It is not difficult to know to order $V^n$ the denominator of 
$\delta_n^n$ is the multiplication of $(cos[\frac {2\pi} {c} (a+b)]-1), 
(cos[\frac {4\pi} {c} (a+b)]-1), ..., (cos[\frac {2n\pi} {c} (a+b)]-1)$. 
We immediately know that $\delta_n$ is always divergent for any 
rational $(a+b)/c$. Unlike the standard FK model, this divergence can 
simply be eliminated by fixing the phase factor to $0$ or $c$ because 
the numerator of $\delta_n^n$ is a sum of a sine and a cosine functions. 
It seems that perturbation theory \cite{cf}  would not work; 
however, this is not the case. In the following, we will show how it 
works perfectly. This divergence only implies there is a new 
commensurate phase which is translationally invariant and unlocked.

First, we group the solutions ..., $\delta_{n-1}, \delta_n, 
\delta_{n+1}$, ..., etc. with two in a group: $( \delta_{n-2}, \delta_{n-
1})_{(i-1)}, ( \delta_n, \delta_{n+1})_i, (\delta_{n+2}, \delta_{n+3})_{(i+1)}$, etc. We 
define a new sequence {$t_i=\delta_n+\delta_{n+1}$}. To first 
order in $\delta_n$, we have $t_i^1=\delta_n^1+\delta_{n+1}^1$. 
From
Eq.~(\ref{psolution1}), we get

\begin{equation}
t_i^1=\frac {(Acos(\frac {\pi a} {c})+Bsin(\frac {\pi a} {c}))sin[\frac 
{2\pi} {c} ((a+b)i+\theta+\frac a2)]} {cos[\frac {2\pi} {c} (a+b)]-1}
\label{lockedeq}
\end{equation}
By setting $\theta+ a/2=0$ or $c$, the divergence can be eliminated 
and $t_i^1$ is locked to be identically zero. However, this does not 
mean the particles are pinned to the external potential. It only shows 
that the two nearest particles move in a different direction, i.e., 
$\delta_n^1=-\delta_{n+1}^1$, etc.. Imposing this condition on 
Eq.~(\ref{pertubationeq2}), we get the finite and nonzero solutions

\begin{equation}
\delta^1_n=\frac {Vsin[\frac {2\pi}{c} (E_n+\theta)]}{2(k_1+k_2)}
\label{elimination}
\end{equation}
From Eq.~(\ref{psolution2}), we can extend the above conclusions to 
second order in $\delta_n$. It is obvious that to any order the 
divergence in $\delta_n$ can be eliminated in this way and we can 
find a finite and nonzero solution like Eq.~(\ref{elimination}). 
Summing over all these terms, we can write the solution to 
Eq.~(\ref{pertubationeq1}) for any rational value of $(a+b)/c$ as

\begin{equation}
\begin{array}{r}
\delta_n=\sum_l \{f_lsin[\frac{2\pi l}{c}(E_n +\theta)]+(-
1)^nt_lcos[\frac{2\pi l}{c}(E_n+\theta)]\}+\\
\sum_m \frac{p_m}{2(k_1+k_2)} sin[\frac {2\pi m}{c} (E_n 
+\theta)]
\end{array}
\label{solution}
\end{equation}
Here $l(a+b)/c\neq integer$, $m(a+b)/c=integer$, and 
$m(\theta+a/2)/c=integer$. $f_l$ is the sum of the solutions to 
Eq.~(\ref{pertubationeq1}) to all orders that contain the function 
$sin[\frac{2\pi l}{c}(E_n +\theta)]$; $t_l$ is the sum of solutions to 
Eq.~(\ref{pertubationeq1}) to all orders that contain the function $ 
cos[\frac{2\pi l}{c}(E_n +\theta)]$. For example, to second order in $V$, 
$f_1=A/[2(cos[\frac {2\pi} {c} (a+b)]-1)]$, $f_2=A_1/[8(cos[\frac 
{2\pi} {c} (a+b)]-1) (cos[\frac {4\pi} {c} (a+b)]-1)]$ and 
$t_1=B/[2(cos[\frac {2\pi} {c} (a+b)]-1)]$, $t_2=B_1/[8(cos[\frac 
{2\pi} {c} (a+b)]-1) (cos[\frac {4\pi} {c} (a+b)]-1)]$. In high orders 
in $V$, they are a very complex sum. $p_m$ is the coefficient of the 
function of $sin[\frac{2\pi m}{c}(E_n +\theta)]$ in 
Eq.~(\ref{pertubationeq1}) just as shown in Eq.~(\ref{elimination}). 
It is easy to show 

\begin{equation}
sin[\frac{2\pi m}{c}(E_n +\theta)]=(-1)^n sin[\frac{2\pi m}{c} \theta 
]
\end{equation}
from the conditions $m(a+b)/c=integer$, and 
$m(\theta+a/2)/c=integer$. So the second sum is independent of the 
the particle's positions; otherwise, this term has different signs for 
the particles at even and odd sites. This shows that all particles with 
$n=even$  move the same distance in the same direction and particles 
with $n=odd$ move in the opposite direction with the same distance. 
In other words, the two sublattices have a relative parallel movement. 
The distance of the movement is dependent on the overall phase 
factor $\theta$. $\theta$ is determined by the condition 
$m(\theta+a/2)/c=integer$. It is related to the parameter $a$. 
However, the parameter $a$ is completely free because the equation 
$b=ga$ does not impose any restriction on it. Therefore, the overall 
phase factor $\theta$ has a lot of freedom in our model. 

Before we go further, let us recall some interesting properties of the 
standard FK model. In the FK model, the perturbative solution 
\cite{cf} for a commensurate phase can be written as 

\begin{equation}
\delta_n=\sum_l h_l sin[\frac {2\pi l}{c} (E_n+\theta)]
\label{fksolution}
\end{equation}
Here $E_n=a_0 n$, $la_0/c\neq integer$. This solution can be viewed 
as a special case of Eq.~(\ref{solution}) at $g=1$. It is an odd 
function. Therefore, a commensurate phase always has a symmetric 
point. Now let us inspect a more  specific case. For $a_0/c$=$ M/N$, 
$\theta/c$ can take the value $n_0/N$ with $n_0=0,...,N$. We have 

\begin{equation}
\frac {E_n+\theta}{c}=\frac {na_0}{c}+\frac{\theta}{c}=\frac 
{nM+n_0}{N}
\label{fixeq}
\end{equation}
It is not difficult to find that there is an infinite set of $n$ values that 
make $(nM+n_0)/c$ equal to an integer for any group of $M$, 
$N$ and $n_0$. This means that there is an infinite set of particles 
that never move with a sinusoidal potential $V$ for any 
commensurate phases and any overall phase factor in the standard FK 
model. They stay at their initial positions for ever no matter how 
$\theta$ is changed. Hence a commensurate phase in the standard FK 
model is always pinned to the substrate and is not translationally 
invariant. 

Now let us return to our model. The most obvious feature of 
Eq.~(\ref{solution}) is that there is no obvious symmetry because it is 
a sum of an odd function and an even function. Thus there is  a symmetry
breaking in our model. A commensurate phase in our 
model does not have any symmetric point. It breaks the symmetry 
of a commensurate phase in the standard FK model. This is similar to 
an incommensurate phase in the FK model. Another interesting 
property is that no particles are pinned to their initial positions. This 
is because there is no way to make a sine function and a cosine 
function to be equal to zero at the same value of their variables. 
Moreover, all particles are related through the overall phase factor 
$\theta$. Its effect cannot be simply eliminated as shown in 
Eq.~(\ref{fixeq}) or otherwise. Any value of $\theta$ has an effect on 
all the particles. Any change in $\theta$ will make all particles move. 
There is no position that a particle is favoured. We have shown that 
$\theta$ has much freedom. Therefore, a commensurate phase in our 
model is translationally invariant and unpinned. However, the 
movement revealed in Eq.~(\ref{solution}) is not trivial. As we have 
shown, the second sum in Eq.~(\ref{solution}) indicates a relative 
parallel movement of the two sublattices. The first sum also shows 
the trend of two sublattices moving in two different directions because 
there is a change of sign for $n=odd$ and $ n=even$. So the basic 
movement in our model is a relatively sliding between two 
sublattices. As a result a commensurate phase is not expected to have 
a zero frequency phonon mode for the whole system because it 
requires the whole system to move in the same direction. So far we 
have shown that our model has a new commensurate phase which is 
translationally invariant and unpinned. 

Except for the new phase, our model also has the usual FK model-
like commensurate phase for the special rational  $\frac {a+b} {c}$. 
When $k_1sin(\frac {2\pi}{c}a)=k_2sin(\frac {2\pi} {c}b), 
k_1sin(\frac {4\pi} {c}a)=k_2sin(\frac {4\pi}{c}b)$, etc., the 
solutions $\delta_n^1, \delta_n^2$, etc. become 

\[
\delta_n^1=\frac {Asin[\frac {2\pi} {c} (E_n+\theta)]} 
{2(cos[\frac {2\pi} {c} (a+b)]-1)} 
\]
\[
\delta_n^2=\frac {A_1sin[\frac {4\pi} {c} (E_n+\theta)]} 
{8(cos[\frac {2\pi} {c} (a+b)]-1) (cos[\frac {4\pi} {c} 
(a+b)]-1)} 
\] 
etc., which are just the same as the solutions \cite{cf} of the FK 
model. The only way to meet these conditions is $a/c$ and $b/c$ are 
the two non-zero and non-negative solutions of the equation $sin(2\pi 
X)=0$, i.e., they are integers or half integers greater than $1/2$. In 
this case, our model behaves like the standard FK model. All the 
conclusions of the FK model can directly be applied to it. These 
commensurate phases are locked and are not translationally 
invariant.

Now we turn to an irrational $(a+b)/ c$. Like the FK model, the 
perturbation theory works well and there is no locking phenomenon 
for any irrational $(a+b)/c$. Therefore, an incommensurate phase of 
our model is translationally invariant and has a zero-frequency 
phonon mode. For a small $k$, $B$, $B_1$, ...$\simeq$0, and $A$ 
and $A_1$, ...$\simeq \frac 2k_1 (1+ cos[\frac {\pi} {c} (a+b)])$ and 
$\frac {8\pi V} {k_1^2c} (1+cos[\frac {\pi} {c} (a+b)])(1+ cos[\frac 
{2\pi} {c} (a+b)])$, etc.. The solutions (12) and (13) become the 
same as the standard FK model \cite{cf}. Aside from this, there is no 
way to make our model behave like the FK model for an irrational 
$(a+b)/c$.

We plot the phase diagram of our model for small $V$ in $a/c$ versus 
$b/c$, as shown in Fig. 2. Phases in the parameter space that form a 
square lattice with the lattice constant $1/2$ are the locked phases 
which are not translationally invariant and do not have a zero-
frequency phonon mode. The others in the phase space 
represent  unlocked phases which are translationally invariant. So a 
phase transition breaking translational symmetry must happen on the 
square lattice in the parameter space $(a/c, b/c)$.


How this phase diagram changes with the amplitude $V$ of the 
external potential is an interesting question. For 
an irrational $(a+b)/c$, no new things are expected to happen: an 
unlocked incommensurate phase is simply locked to a translationally 
invariant incommensurate phase. However, for an unlocked 
commensurate phase, a new phase transition is expected to occur. It 
will definitely be locked to the external potential with the increase of 
$V$. It is very possible that an unlocked commensurate phase is 
locked to a phase on the square lattice in the parameter space. If this 
happens, the phase transition would break translational symmetry.

In summary, we have proposed a new generalised FK model. Its 
continuum version is different from the sine-Gordon equation. Under 
certain conditions, our model is the same as the discrete sine-Gordon 
equation.  Surprisingly, it occurs at a large difference of $(a-b)/c > 1/2$ 
instead of a small one for a rational $(a+b)/c$. 
More interestingly, our model has a translationally invariant 
commensurate phase which is unpinned. It has a rich phase diagram, 
which is to be explored in the future. A phase transition breaking 
translational symmetry take place on a square lattice in the parameter 
space. Furthermore, we predict another new phase transition would 
happen for large $V$.

\vspace{2cm}
{\parindent=0pt {\Large{\bf{Acknowledgements}}}}

This work is supported in part by the Research Grant Council 
RGC/96-97/10 and the Hong Kong Baptist University Faculty 
Research Grant FRG/95-96/II-09 and FRG/95-96/II-92.


\newpage

{\Large{\bf{Figure Captions:}}}
\begin{enumerate}
\item Frenkel-Kontorova model with two spring constants $k_1$ and $k_2$.

\item Phase diagram $a/c$ vs. $b/c$. A dot represents a locked phase; the 
others represent unpinned phases.
\end{enumerate}


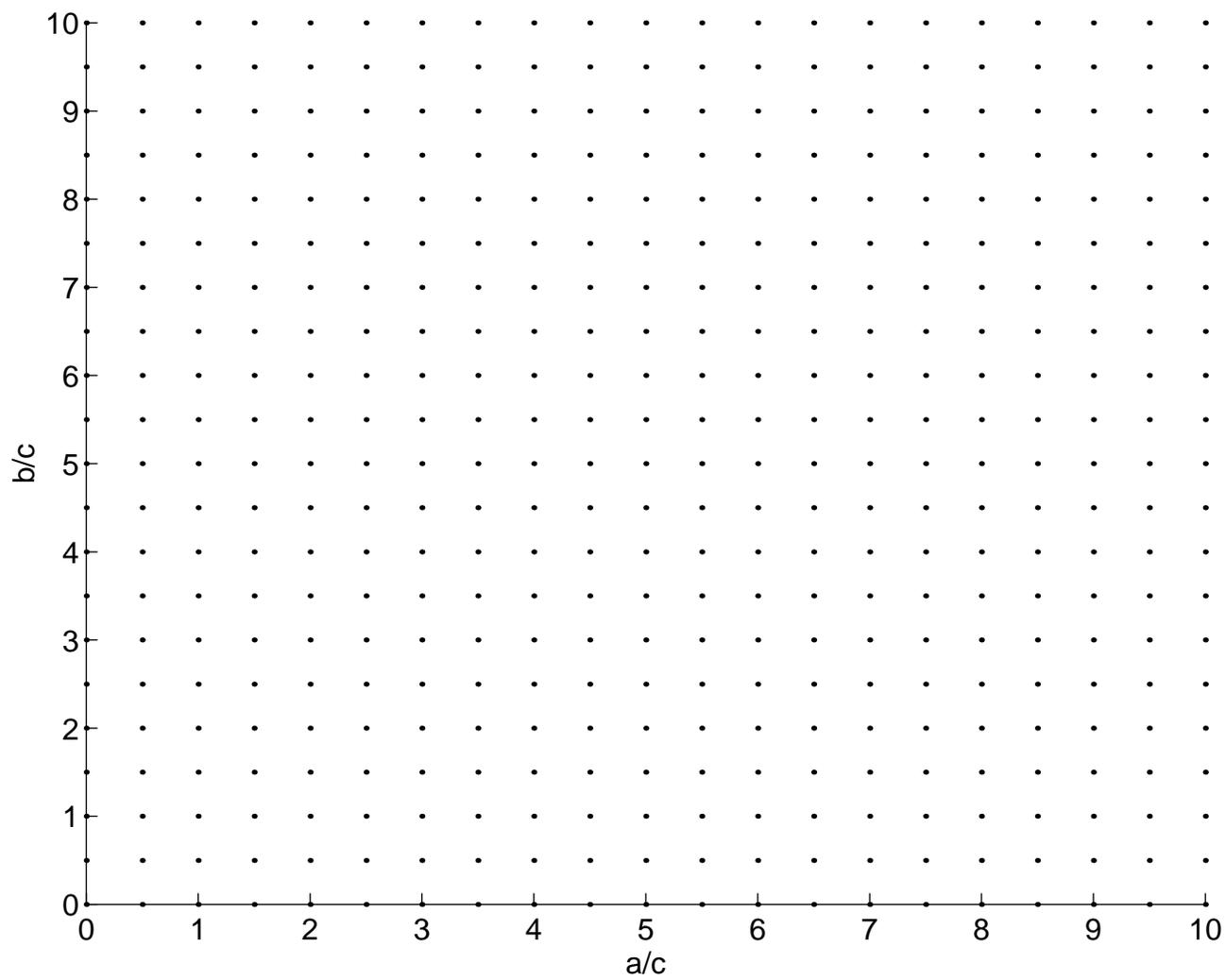
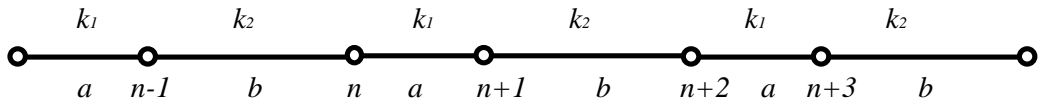
\end{document}